\documentclass[12pt]{article}
\usepackage{graphicx}
\usepackage[cp1251]{inputenc}

\begin{document}
 \noindent {\footnotesize\it Astronomy Letters, 2019, Vol. 45, No 6, pp. 331--340.}
 \newcommand{\dif}{\textrm{d}}

 \noindent
 \begin{tabular}{llllllllllllllllllllllllllllllllllllllllllllll}
 & & & & & & & & & & & & & & & & & & & & & & & & & & & & & & & & & & & & & \\\hline\hline
 \end{tabular}

  \vskip 0.5cm
  \centerline{\bf\large Galactic Rotation Based on OB Stars from}
  \centerline{\bf\large the Gaia DR2 Catalogue}
   \bigskip
  \bigskip
  \centerline
 {V.V. Bobylev\footnote [1]{e-mail: vbobylev@gaoran.ru} and A.T. Bajkova}
  \bigskip

  \centerline{\small\it Pulkovo Astronomical Observatory, Russian Academy of Sciences,}

  \centerline{\small\it Pulkovskoe sh. 65, St. Petersburg, 196140 Russia}
 \bigskip
 \bigskip
 \bigskip

 {
{\bf Abstract}---We have studied a sample containing $\sim­6000$
OB stars with proper motions and trigonometric parallaxes from the
Gaia DR2 catalogue. The following parameters of the angular
velocity of Galactic rotation have been found:
 $\Omega_0=29.70\pm0.11$ km s$^{-1}$ kpc$^{-1}$,
 $\Omega'_0=-4.035\pm0.031$ km s$^{-1}$ kpc$^{-2}$, and
 $\Omega''_0= 0.620\pm0.014$ km s$^{-1}$ kpc$^{-3}$.
The circular rotation velocity of the solar neighborhood around
the Galactic center is $V_0=238\pm5$ km s$^{-1}$ for the adopted
Galactocentric distance of the Sun $R_0=8.0\pm0.15$ kpc. The
amplitudes of the tangential and radial velocity perturbations
produced by the spiral density wave are $f_\theta=4.4\pm1.4$ km
s$^{-1}$ and $f_R=5.1\pm1.2$ km s$^{-1}$, respectively; the
perturbation wavelengths are $\lambda_\theta=1.9\pm0.5$ kpc and
$\lambda_R=2.1\pm0.5$ kpc for the adopted four-armed spiral
pattern. The Sun's phase in the spiral density wave is
$\chi_\odot=-178^\circ\pm12^\circ$.
  }


 \subsection*{INTRODUCTION}
Stars of spectral types O and B are an important tool for studying
the Galaxy and its subsystems. Being young, they trace well the
spiral arms, because most of them have not receded very far
(except for the small percentage of high-velocity runaway stars)
from their birthplace. OB stars have nearly circular Galactic
orbits and, therefore, serve as an excellent material for studying
the Galactic rotation. They are visible from great distances and,
therefore, are well suited for studying the Galactic disk
structure, the spiral pattern, the central bar, young open star
clusters, OB associations, and various star-forming regions. Such
studies were performed, for example, by Y.M. Georgelin and Y.P.
Georgelin (1976), Byl and Ovenden (1978), Maiz-Apell\'aniz (2001),
Zabolotskikh et al. (2002), and Russeil (2003).

At present, the Gaia space experiment (Brown et al. 2016) is a
valuable data source for studying the structure and kinematics of
the Galaxy. The Gaia second data release, Gaia DR2, was published
in April 2018 (Brown et al. 2018; Lindegren et al. 2018). It
contains the trigonometric parallaxes and proper motions of
$\sim$1.3 billion stars. The mean trigonometric parallax errors
lie in the range 0.02--0.04 mas for bright stars ($G<15^m$) and
reach 0.7 mas for faint stars ($G=20^m$). For more than 7 million
stars of spectral types F--G--K the line-of-sight velocities have
been determined with a mean error of $\sim$1 km s$^{-1}$. The
errors in the line-of-sight velocities of OB stars are known to be
considerably larger due to the peculiarities of the spectra for
these stars.

Highly accurate Gaia DR2 data have already yielded a number of
significant kinematic results. For example, based on Gaia DR2
data, Helmi et al. (2018) determined new proper motions for 75
Galactic globular clusters and a number of dwarf galaxies that are
Milky Way satellites, including the Magellanic Clouds. A new
estimate of the Galactic mass, $9.8^{+6.7}_{-2.7}\times10^{11}
M_\odot$, was obtained by analyzing their space velocities. Based
on a kinematic analysis of the space velocities from the Gaia DR2
catalogue, Antoja et al. (2018) detected some manifestations of
the still ongoing wobble of the Milky Way disk after the passage
of the Sagittarius dwarf galaxy through it.

Based on data from the Gaia DR2 catalogue, Cantat-Gaudin et al.
(2018) determined new mean proper motions for 1229 open star
clusters (OSCs), while for a significant fraction of this OSC list
Soubiran et al. (2018) deduced the mean values of their
line-of-sight velocities exclusively from Gaia DR2 data. The
kinematics of these OSCs was studied by Bobylev and Bajkova
(2019). The spatial and intrinsic kinematic properties of a number
of young stellar associations close to the Sun have been studied
with hitherto unprecedented detail (Zari et al. 2018; Franciosini
et al. 2018; Roccatagliata et al. 2018; Kounkel et al. 2018).

A sample containing $\sim$500 OB stars with parallaxes and proper
motions from the Gaia DR2 catalogue was studied by Bobylev and
Bajkova (2018b). The goal of this paper is to improve the Galactic
rotation parameters and the parameters of the Galactic spiral
density wave based on a huge sample of OB stars for which the
trigonometric parallaxes and proper motions are available in the
Gaia DR2 catalogue. Such a sample of OB stars containing data on
5772 stars has recently been published by Xu et al. (2018).

 \section*{METHOD}
 \subsection*{Galactic Rotation Parameters}
We know three stellar velocity components from observations: the
line-of-sight velocity $V_r$ and the two tangential velocity
components $V_l=4.74r\mu_l\cos b$ and $V_b=4.74r\mu_b$ along the
Galactic longitude $l$ and latitude $b,$ respectively, expressed
in km s$^{-1}$. Here, the coefficient 4.74 is the ratio of the
number of kilometers in an astronomical unit to the number of
seconds in a tropical year, and $r=1/\pi$ is the stellar
heliocentric distance in kpc that we calculate via the stellar
parallax $\pi.$ The proper motion components $\mu_l\cos b$ and
$\mu_b$ are expressed in mas yr$^{-1}$.

To determine the parameters of the Galactic rotation curve, we use
the equations derived from Bottlinger’s formulas, in which the
angular velocity $\Omega$ is expanded into a series to terms of
the second order of smallness in $r/R_0$:
\begin{equation}
 \begin{array}{lll}
 V_r=-U_\odot\cos b\cos l-V_\odot\cos b\sin l-W_\odot\sin b\\
 +R_0(R-R_0)\sin l\cos b\Omega^\prime_0
 +0.5R_0(R-R_0)^2\sin l\cos b\Omega^{\prime\prime}_0,
 \label{EQ-1}
 \end{array}
 \end{equation}
 \begin{equation}
 \begin{array}{lll}
 V_l= U_\odot\sin l-V_\odot\cos l-r\Omega_0\cos b\\
 +(R-R_0)(R_0\cos l-r\cos b)\Omega^\prime_0
 +0.5(R-R_0)^2(R_0\cos l-r\cos b)\Omega^{\prime\prime}_0,
 \label{EQ-2}
 \end{array}
 \end{equation}
where $R$ is the distance from the star to the Galactic rotation
axis:
  \begin{equation}
 R^2=r^2\cos^2 b-2R_0 r\cos b\cos l+R^2_0.
 \end{equation}
The quantity $\Omega_0$ is the angular velocity of Galactic
rotation at the solar distance $R_0,$ the parameters
$\Omega^{\prime}_0$ and $\Omega^{\prime\prime}_0$ are the
corresponding derivatives of this angular velocity, and the linear
rotation velocity of the Galaxy is $V_0=|R_0\Omega_0|.$ In Eqs.
(1) and (2) the five unknowns $U_\odot, V_\odot, \Omega_0,
\Omega^{\prime}_0$ and $\Omega^{\prime\prime}_0$ are to be
determined; in Eq. (1) there are only four unknown parameters,
because the angular velocity of rotation $\Omega_0$ is absent.

Note a number of studies devoted to determining the mean distance
from the Sun to the Galactic center using its individual
determinations in the last decade by independent methods. For
example, $R_0=8\pm0.2$ kpc (Vall\'ee 2017), $R_0=8.4\pm0.4$ kpc
(de Grijs and Bono 2017), or $R_0=8\pm0.15$ kpc (Camarillo et al.
2018). Based on these reviews, here we adopted $R_0=8\pm0.15$ kpc.

The kinematic parameters are determined by solving the conditional
equations (1) and (2) by the least-squares method (LSM). We use
weights of the form $w_r=S_0/\sqrt {S_0^2+\sigma^2_{V_r}}$ and
$w_l=S_0/\sqrt {S_0^2+\sigma^2_{V_l}},$ where $S_0$ is the
``cosmic'' dispersion, $\sigma_{V_r}$ and $\sigma_{V_l}$ are the
dispersions of the corresponding observed velocities. $S_0$ is
comparable to the root-mean-square residual $\sigma_0$ (the error
per unit weight) that is calculated when solving the conditional
equations (1) and (2). In this paper we adopted $S_0=12$ km
s$^{-1}$ typical for young stars. The system of equations (1) and
(2) is solved in several iterations using the $3\sigma$ criterion
to eliminate the stars with large residuals.

When the stars distributed uniformly over the celestial sphere are
analyzed, a third equation, where the components $V_b$ are on the
left-hand side, is also used. When analyzing the OB stars located
virtually in the Galactic plane, where $\sin b\approx0,$ using the
equation with $V_b$ is ineffective. However, even when the system
of conditional equations (1) and (2) is solved simultaneously, the
velocity $W_\odot$ is determined poorly. Therefore, in this paper
we assume it to be known, $W_\odot=7$ km s$^{-1}$.

 \subsection*{Influence of the Spiral Density Wave}
To study the influence of the Galactic spiral density wave, it is
first necessary to calculate three space velocities $U,V,W$ for
each star, where $U$ is directed from the Sun to the Galactic
center, $V$ is in the direction of Galactic rotation, and $W$ is
directed to the north Galactic pole. These velocities are
calculated via the components $V_r,$ $V_l,$ and $V_b$:
 \begin{equation}
 \begin{array}{lll}
 U=V_r\cos l\cos b-V_l\sin l-V_b\cos l\sin b,\\
 V=V_r\sin l\cos b+V_l\cos l-V_b\sin l\sin b,\\
 W=V_r\sin b                +V_b\cos b.
 \label{UVW}
 \end{array}
 \end{equation}
Thus, they can be determined only for those stars for which both
line-of-sight velocities and proper motions have been measured.

We can find two velocities, $V_R$ directed radially away from the
Galactic center and the velocity $V_{circ}$ orthogonal to it
pointing in the direction of Galactic rotation, based on the
following relations:
 \begin{equation}
 \begin{array}{lll}
  V_{circ}= U\sin \theta+(V_0+V)\cos \theta, \\
       V_R=-U\cos \theta+(V_0+V)\sin \theta,
 \label{VRVT}
 \end{array}
 \end{equation}
where the position angle $\theta$ obeys the relation
$\tan\theta=y/(R_0-x)$, and $x,y,z$ are the rectangular
heliocentric coordinates of the star (the velocities $U,V,W$ are
directed along the corresponding $x,y,z$ axes), $V_0$ is the
linear rotation velocity of the Galaxy at the solar distance
$R_0.$ The velocities $V_R$ and $W$ are virtually independent of
the pattern of the Galactic rotation curve. To analyze the
periodicities in the tangential velocities, it is necessary to
form the residual velocities $\Delta V_{circ}$ by taking into
account a smoothed Galactic rotation curve.

According to the linear theory of density waves (Lin and Shu
1964), the influence of the spiral density wave is described by
the following relations:
 \begin{equation}
 \begin{array}{lll}
       V_R =-f_R \cos \chi,\\
 \Delta V_{circ}= f_\theta \sin\chi,
 \label{DelVRot}
 \end{array}
 \end{equation}
where
 \begin{equation}
 \chi=m[\cot(i)\ln(R/R_0)-\theta]+\chi_\odot
 \end{equation}
is the phase of the spiral density wave ($m$ is the number of
spiral arms, $i$ is the pitch angle of the spiral pattern, and
$\chi_\odot$ is the Sun’s radial phase in the spiral density
wave); $f_R$ and $f_\theta$ are the amplitudes of the radial and
tangential velocity perturbations, which are assumed to be
positive. As an analysis of the present day highly accurate data
showed, the periodicities associated with the spiral density wave
also manifest themselves in the vertical velocities $W$ (Bobylev
and Bajkova 2015; Rastorguev et al. 2017).

We apply a modified spectral analysis (Bajkova and Bobylev 2012)
to study the periodicities in the velocities $V_R$ and $\Delta
V_{circ}$. The wavelength $\lambda$ (the distance between adjacent
spiral arm segments measured along the radial direction) is
calculated from the relation
\begin{equation}
 \lambda m \cot(i)=2\pi R_0.
 \label{a-04}
\end{equation}
Let there be a series of measured velocities $V_{R_n}$ (these can
be both radial ($V_R$) and tangential ($\Delta V_{circ}$)
velocities), $n=1,\dots,N$, where $N$ is the number of objects.
The objective of our spectral analysis is to extract a periodicity
from the data series in accordance with the adopted model
describing a spiral density wave with parameters  $f,$
$\lambda$~(or $i)$ and $\chi_\odot$.

Having taken into account the logarithmic behavior of the spiral
density wave and the position angles of the objects $\theta_n$,
our spectral (periodogram) analysis of the series of velocity
perturbations is reduced to calculating the square of the
amplitude (power spectrum) of the standard Fourier transform
(Bajkova and Bobylev 2012):
\begin{equation}
 \bar{V}_{\lambda_k} = \frac{1} {N}\sum_{n=1}^{N} V^{'}_n(R^{'}_n)
 \exp\biggl(-j\frac {2\pi R^{'}_n}{\lambda_k}\biggr),
 \label{29}
\end{equation}
where $\bar{V}_{\lambda_k}$ is the $k$th harmonic of the Fourier
transform with wavelength $\lambda_k=D/k$, $D$ is the period of
the series being analyzed,
 \begin{equation}
 \begin{array}{lll}
 R^{'}_{n}=R_0\ln(R_n/R_0),\\
 V^{'}_n(R^{'}_n)=V_n(R^{'}_n)\times\exp(jm\theta_n).
 \label{21}
 \end{array}
\end{equation}
The sought-for wavelength $\lambda$ corresponds to the peak value
of the power spectrum $S_{peak}$. The pitch angle of the spiral
density wave is derived from Eq. (8). The pitch angle of the
spiral density wave is derived from Eq. (9). We determine the
perturbation amplitude and phase by fitting the harmonic with the
wavelength found to the observational data. The following relation
can also be used to estimate the perturbation amplitude:
 \begin{equation}
 f_R(f_\theta)=2\times\sqrt{S_{peak}}.
 \label{Speak}
 \end{equation}
Thus, our approach consists of two steps: (i) the construction of
a smooth Galactic rotation curve and (ii) a spectral analysis of
the radial ($V_R$) and residual tangential ($\Delta V_{circ}$)
velocities. This method was applied by Bobylev and Bajkova (2012,
2018b) to study the kinematics of young Galactic objects.

 \subsection*{Monte Carlo Simulations}
We use Monte Carlo simulations to estimate the errors in the
parameters of the spiral density wave being determined. In
accordance with this method, we generate $M$ independent
realizations of data on the parallaxes and velocities of objects
with their random measurement errors that are known to us.

We assume that the measurement errors of the data are distributed
normally with a mean equal to the nominal value and a dispersion
equal to $\sigma_l={error}_l, l=1,\dots,N_d$, where $N_d$ is the
number of data and $error_l$ denotes the measurement error of a
single measurement with number $l$ (one sigma). Each element of a
random realization is formed independently by adding the nominal
value of the measured data with number $l$ and the random number
generated according to a normal law with zero mean and dispersion
$\sigma_l$. Note that the latter is limited from above by
$3\sigma_l$.

Each random realization of data with number $j$ ($j=1,\dots,M$)
generated in this way is then processed according to the algorithm
described above to determine the sought-for parameters $f_R^j,
\lambda^j,$ and $\chi_\odot^j$. The mean values of the parameters
and their dispersions are then determined from the derived
sequences of estimates: $m_{f_R}\pm \sigma_{f_R}, m_{\lambda}\pm
\sigma_{\lambda}, m_{\chi_\odot}\pm \sigma_{\chi_\odot}$. The
statistical parameters of the spiral density wave pitch angle $i$
can be determined using Eq. (8): $m_{i}\pm\sigma_{i}$.

 \section*{DATA}
In this paper we use the catalogue of OB stars produced by Xu et
al. (2018). It gives the proper motions and trigonometric
parallaxes taken from the Gaia DR2 catalogue for 5772 stars of
spectral types O--B2. For $\sim$2000 OB stars it gives the
line-of-sight velocities taken from the
SIMBAD\footnote{http://simbad.u-strasbg.fr/simbad/} electronic
database. Note that in the catalogue by Xu et al. (2018) the
line-of-sight velocities of OB stars are given relative to the
local standard of rest and, therefore, we convert them to the
heliocentric ones in advance using the parameters of the standard
solar apex $(U,V,W)_\odot=(10.3,15.3,7.7)$ km s$^{-1}$.

 \subsection*{Correction to the Gaia DR2 Parallaxes}
The presence of a possible systematic offset $\Delta\pi=-0.029$
mas in the Gaia DR2 parallaxes with respect to an inertial
reference frame was first pointed out by Lindegren et al. (2018).
Here, the minus means that this correction should be added to the
Gaia DR2 stellar parallaxes to reduce them to the standard.

Arenou et al. (2018) provided an overview of the results of
comparing the Gaia DR2 parallaxes with 29 independent distance
scales that confirm the presence of an offset in the Gaia DR2
parallaxes $\Delta\pi\sim-0.03$ mas. The discrepancies between
individual results turned out to be very large (stars of the
Hipparcos and RECONS programs, stars of the dwarf galaxies
Phoenix, Leo I, and Leo II); as a result, Arenou et al. (2018) did
not deduce the mean value for this correction. Subsequently, some
of the results used by Arenou et al. (2018) were confirmed by
other authors based on new data. For example, based on RR Lyrae
stars, Muraveva et al. (2018) found the correction
$\Delta\pi=-0.057$ mas, or based on 88 radio stars whose
trigonometric parallaxes were measured by various authors by means
of VLBI, Bobylev (2019) obtained an estimate of
$\Delta\pi=-0.038\pm0.046$ mas.

Completely new results are of greatest interest. Stassun and
Torres (2018) found the correction $\Delta\pi=-0.082\pm0.033$ mas
by comparing the parallaxes of 89 detached eclipsing binaries with
their trigonometric parallaxes from the Gaia DR2 catalogue. These
stars were selected from published data using very rigorous
criteria imposed on the photometric parameters. As a result, the
relative errors in the stellar radii, effective temperatures, and
bolometric luminosities, from which the distances are estimated,
do not exceed 3\%.

By comparing the Gaia DR2 trigonometric parallaxes and photometric
parallaxes of 94 OSCs, Yalyalieva et al. (2018) found the
correction $\Delta\pi=-0.045\pm0.009$ mas. The high accuracy of
this estimate is attributable to the high accuracy of the
photometric distance estimates for OSCs obtained by invoking
first-class infrared photometric surveys, such as IPHAS, 2MASS,
WISE, and Pan-STARRS.

Riess et al. (2018) obtained an estimate of
$\Delta\pi=-0.046\pm0.013$ mas based on a sample of 50 long period
Cepheids when comparing their parallaxes with those from the Gaia
DR2 catalogue. The photometric parameters of these Cepheids
measured from the Hubble Space Telescope were used.

By comparing the distances of $\sim$3000 stars from the APOKAS-2
catalogue (Pinsonneault et al. 2018) belonging to the red giant
branch with the Gaia DR2 data, Zinn et al. (2018) found the
correction $\Delta\pi=-0.053\pm0.003$ mas. These authors also
obtained a close value, $\Delta\pi=-0.050\pm0.004$ mas, by
analyzing stars belonging to the red giant clump. The distances to
such stars were estimated from asteroseismic data. According to
these authors, the parallax errors here are approximately equal to
the errors in estimating the stellar radius and are, on average,
1.5\%. Such small errors in combination with the enormous number
of stars allowed $\Delta\pi$ to be determined with a high
accuracy.

The listed results lead to the conclusion that the trigonometric
parallaxes of stars from the Gaia DR2 catalogue should be
corrected by applying a small correction. We will be oriented to
the results of Yalyalieva et al. (2018), Riess et al. (2018), and
Zinn et al. (2018), which look most reliable.

 \begin{table}[t]
 \caption[]{\small
The Galactic rotation parameters found from OB stars with proper
motions and trigonometric parallaxes ($\sigma_\pi/\pi<15\%$) for
various corrections to the Gaia DR2 parallaxes
 }
  \begin{center}  \label{t:01}
  \small
  \begin{tabular}{|l|r|r|r|r|r|}\hline
   Parameters                  & $\pi=\pi+0$ mas & $\pi=\pi+0.02$ mas & $\pi=\pi+0.03$ mas & $\pi=\pi+0.05$ mas \\\hline

    $U_\odot,$    km s$^{-1}$            &  $ 9.49\pm0.72$  &  $ 9.19\pm0.71$  &  $  9.24\pm0.70$ & $  9.27\pm0.69$  \\
    $V_\odot,$    km s$^{-1}$            &  $12.59\pm0.69$  &  $13.16\pm0.68$  &  $ 12.81\pm0.69$ & $ 13.24\pm0.67$  \\
  $\Omega^{'}_0,$ km s$^{-1}$ kpc$^{-2}$ & $-3.380\pm0.100$ & $-3.559\pm0.098$ & $-3.591\pm0.097$ & $-3.811\pm0.096$ \\
 $\Omega^{''}_0,$ km s$^{-1}$ kpc$^{-3}$ & $ 0.483\pm0.118$ & $ 0.676\pm0.014$ & $ 0.613\pm0.125$ & $ 0.742\pm0.120$ \\
   $\sigma_0,$    km s$^{-1}$            &            18.32 &            18.35 &           18.54  &           18.54  \\
     $N_\star$                           &             1783 &             1839 &            1876  &            1925  \\
 \hline
    $U_\odot,$    km s$^{-1}$            &   $ 6.12\pm0.26$ & $ 6.25\pm0.25$   & $  6.22\pm0.24$  & $  6.17\pm0.23$  \\
    $V_\odot,$    km s$^{-1}$            &   $ 6.77\pm0.44$ & $ 7.10\pm0.42$   & $  7.13\pm0.40$  & $  7.38\pm0.38$  \\
  $\Omega_0,$     km s$^{-1}$ kpc$^{-1}$ & $29.90\pm0.13$   & $29.76\pm0.13$   & $ 29.73\pm0.13$  & $ 29.56\pm0.12$  \\
  $\Omega^{'}_0,$ km s$^{-1}$ kpc$^{-2}$ & $-4.202\pm0.037$ & $-4.175\pm0.036$ & $-4.154\pm0.035$ & $-4.119\pm0.035$ \\
 $\Omega^{''}_0,$ km s$^{-1}$ kpc$^{-3}$ & $ 0.647\pm0.032$ & $ 0.658\pm0.032$ & $ 0.645\pm0.029$ & $ 0.648\pm0.029$ \\
   $\sigma_0,$    km s$^{-1}$            &           12.17  &           11.84  &           11.73  &           11.31  \\
     $N_\star$                           &            4249  &            4408  &            4488  &            4620  \\
 $(\Omega^{'}_0)_{V_r}/(\Omega^{'}_0)_{V_l}$ & 0.80 &             0.85 &             0.86 &             0.93  \\
  \hline
 \end{tabular}\end{center}
 {\small The results obtained only from the line-of-sight velocities
$V_r$ (Eq. (1)) and only from the component $V_l$ (Eq. (2)) are
given in the upper and lower parts, respectively. $N_\star$ is the
number of stars used.}
 \end{table}
 \begin{table}[t]
 \caption[]{\small
The Galactic rotation parameters found from OB stars with proper
motions and trigonometric parallaxes when simultaneously solving
Eqs. (1) and (2) for various constraints on the relative errors of
the parallaxes from the Gaia DR2 catalogue; the correction
$\Delta\pi=0.050$ mas was added to the parallaxes of OB stars
 }
  \begin{center}  \label{t:02}
  \small
  \begin{tabular}{|l|r|r|r|r|r|}\hline
   Parameters                            & $\sigma_\pi/\pi<10\%$ & $\sigma_\pi/\pi<15\%$ & $\sigma_\pi/\pi<20\%$ & $\sigma_\pi/\pi<30\%$  \\\hline
    $U_\odot,$    km s$^{-1}$            &  $ 6.13\pm0.26$  &  $ 6.48\pm0.24$  & $  6.61\pm0.23$  & $  6.62\pm0.23$  \\
    $V_\odot,$    km s$^{-1}$            &  $10.20\pm0.38$  &  $ 9.95\pm0.33$  & $  9.79\pm0.31$  & $  8.99\pm0.28$  \\
  $\Omega_0,$     km s$^{-1}$ kpc$^{-1}$ & $29.09\pm0.17$   & $29.13\pm0.13$   & $ 29.03\pm0.13$  & $ 29.15\pm0.13$  \\
  $\Omega^{'}_0,$ km s$^{-1}$ kpc$^{-2}$ & $-4.161\pm0.045$ & $-4.058\pm0.035$ & $-3.992\pm0.033$ & $-3.915\pm0.032$ \\
 $\Omega^{''}_0,$ km s$^{-1}$ kpc$^{-3}$ & $ 0.824\pm0.055$ & $ 0.767\pm0.030$ & $ 0.742\pm0.023$ & $ 0.620\pm0.015$ \\
   $\sigma_0,$    km s$^{-1}$            &           12.84  &           13.54  &           13.93  &          14.28  \\
     $N_\star$                           &            3313  &            4569  &            4959  &           5175  \\
             $A,$ km s$^{-1}$ kpc$^{-1}$ &  $ 16.64\pm0.18$ &  $ 16.23\pm0.14$ &  $ 15.97\pm0.13$ & $ 15.66\pm0.13$ \\
             $B,$ km s$^{-1}$ kpc$^{-1}$ &  $-12.45\pm0.24$ &  $-12.89\pm0.19$ &  $-13.06\pm0.18$ & $-13.50\pm0.18$ \\
           $V_0,$ km s$^{-1}$            &  $  232.8\pm4.6$ &  $  233.0\pm4.5$ &  $  232.3\pm4.5$ & $  233.2\pm4.5$ \\
  \hline
 \end{tabular}
 \end{center}
 {\small $N_\star$ is the number of stars used.}
 \end{table}
\begin{figure}[t]
{\begin{center}
  \includegraphics[width=0.55\textwidth]{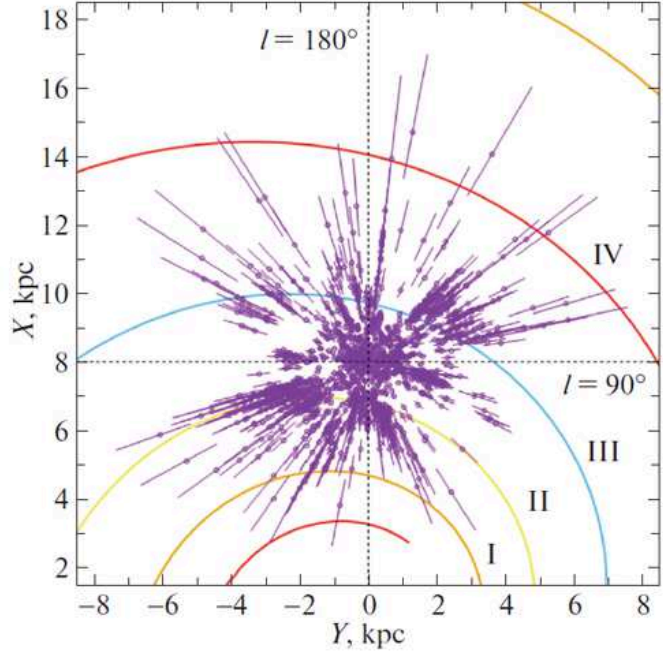}
 \caption{
(Color online) Distribution of the sample of OB stars for which
both line-of-sight velocities and proper motions are available on
the $XY$ plane; the Sun has coordinates $(X,Y)=(8,0)$ kpc, the
four-armed spiral pattern with a pitch angle of $-13^\circ$ is
shown (Bobylev and Bajkova 2014), the spiral arm segments are
numbered by Roman numerals.
  } \label{f-XY}
\end{center}}
\end{figure}
\begin{figure}[t]
{\begin{center}
   \includegraphics[width=0.9\textwidth]{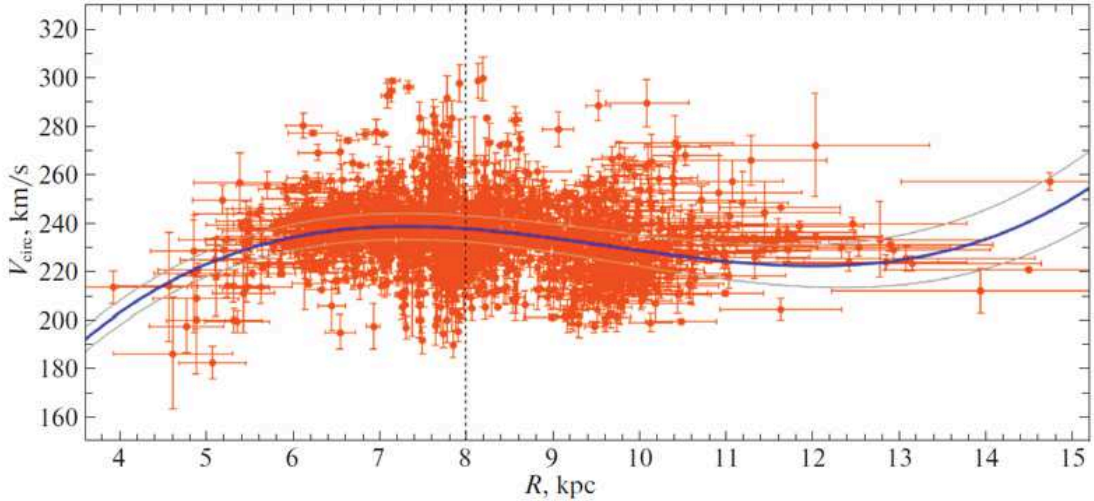}
 \caption{
(Color online) Circular velocities of OB stars versus
Galactocentric distance, the Galactic rotation curve constructed
according to the solution (12) with the $1\sigma$ boundaries of
the confidence intervals; the vertical dashed line marks the Sun's
position.
  } \label{f-Rotat}
\end{center}}
\end{figure}
\begin{figure}[t]
{\begin{center}
  \includegraphics[width=0.55\textwidth]{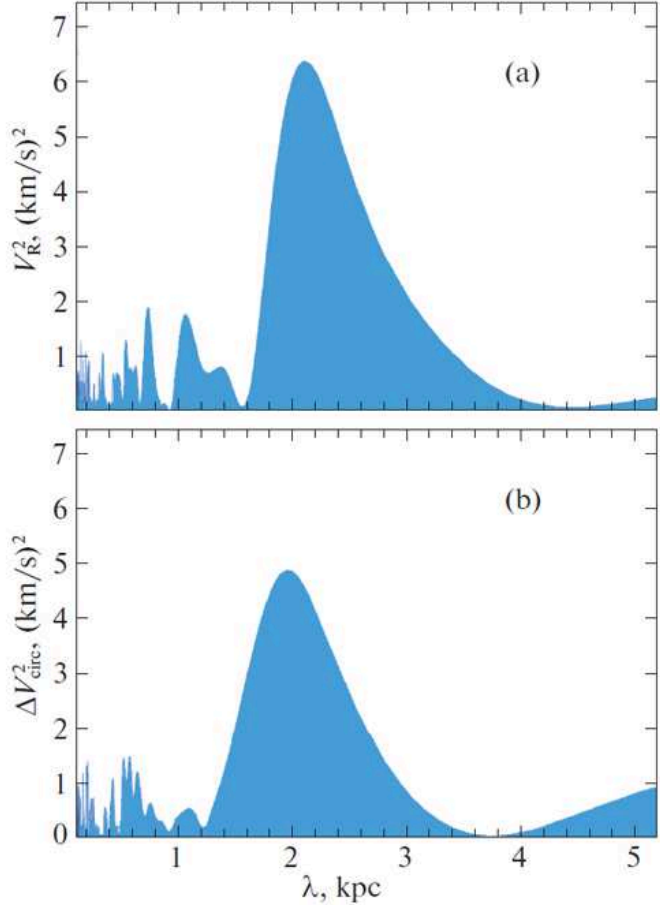}
 \caption{
(Color online) Power spectra for the radial (a) and residual
tangential (b) velocities of OB stars.
  } \label{f-Spectr}
\end{center}}
\end{figure}
\begin{figure}[t]
{\begin{center}
  \includegraphics[width=0.85\textwidth]{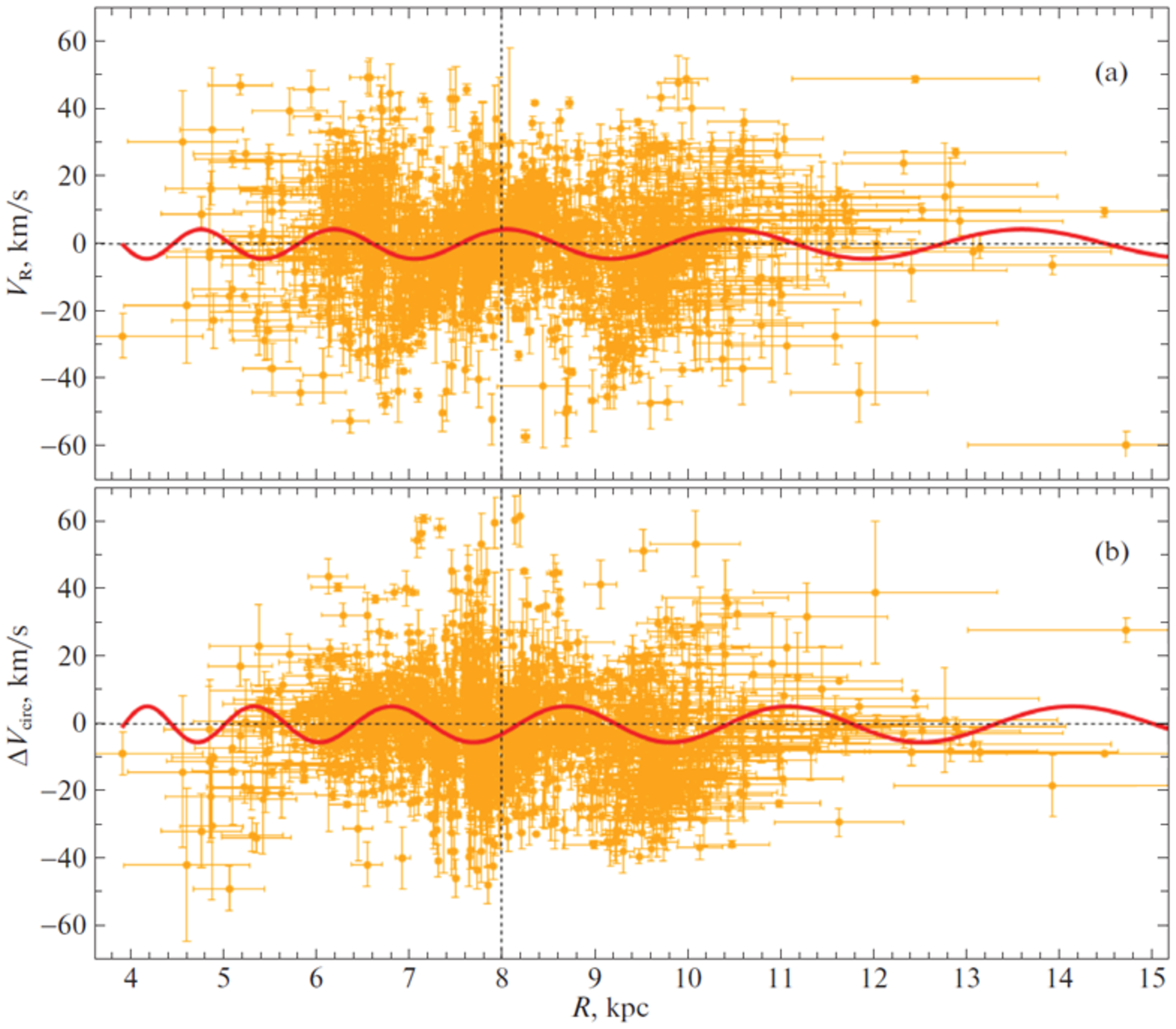}
 \caption{
(Color online) Radial (a) and residual tangential (b) velocities
of OB stars versus Galactocentric distance; the vertical dashed
line marks the Sun's position.
  } \label{f-Rest}
\end{center}}
\end{figure}

 \section*{RESULTS}
 \subsection*{The Entire Sample of OB Stars}
Let us first consider all our OB stars for various constraints on
the relative trigonometric parallax errors and various corrections
of the zero point of the Gaia DR2 parallaxes, $\Delta\pi$.

Table 1 gives the kinematic parameters obtained for various
corrections to the Gaia DR2 parallaxes as a result of the separate
solutions of Eqs. (1) and (2). We used OB stars with relative
trigonometric parallax errors less than 15\%. As can be seen from
the table, there is a noticeable influence of the correction
$\Delta\pi$ when the kinematic parameters are determined using the
line-of-sight velocities of stars (the upper part of Table 1). The
second derivative of the angular velocity of Galactic rotation
$\Omega^{''}_0.$ is affected most strongly here.

The last row in Table 1 gives the ratio of the first derivative of
the angular velocity found using only the line-of-sight velocities
$(\Omega^{'}_0)_{V_r}$ to the one found using only the proper
motions $(\Omega^{'}_0)_{V_l}$.

This method is based on the fact that the estimate of the first
derivative of the angular velocity obtained from the proper
motions depends very weakly on the error of the adopted distance
scale, while the estimate of the first derivative of the angular
velocity obtained from the line-of-sight velocities is inversely
proportional to the adopted scale of the distance scale.
Therefore, comparing the values of $\Omega_0'$ found by various
methods allows the correction factor of the distance scale $p$ to
be found (Zabolotskikh et al. 2002; Rastorguev et al. 2017); in
our case, $p=(\Omega^{'}_0)_{V_r}/(\Omega^{'}_0)_{V_l}$. The error
in $p$ was calculated based on the relation
 $\sigma^2_p=(\sigma_{\Omega'_{0V_r}}/\Omega'_{0V_l})^2+
     (\Omega'_{0V_r}\cdot\sigma_{\Omega'_{0V_l}}/\Omega'^2_{0V_l})^2.$
Having analyzed more than 50 000 stars from the TGAS catalogue
(Brown et al. 2016), Bobylev and Bajkova (2018a) obtained an
estimate of $p=0.97\pm0.04.$ by this method. According to the
results in Table 1, we can see that the distance scale factor $p$
tends to unity as the correction $\Delta\pi$ increases. On the
other hand, the noticeable deviations of this factor from unity in
the first columns of the table are determined almost entirely by
$\Omega'_{0V_r}$. It can be concluded that the quality of the
line-of-sight velocities for this sample of OB stars is low.

Table 2 gives the kinematic parameters found when simultaneously
solving the system of equations (1) and (2) for various
constraints on the relative errors of the parallaxes from the Gaia
DR2 catalogue. Here, the correction $\Delta\pi=0.050$ mas was
added to the parallaxes of OB stars. The simultaneous solution is
as follows. The OB stars with proper motions, line-of-sight
velocities, and distances give two equations, (1) and (2), while
the stars for which only the proper motions are available give
only Eq. (2). As can be seen from Table 1, the stars with
line-of-sight velocities are approximately half those with proper
motions.

Based on a sample of 5335 stars, to the parallaxes of which we
added the correction $\Delta\pi=0.050$ mas, we found the following
kinematic parameters from the solution of only Eq. (2):
 \begin{equation}
 \label{solution-best-Vl}
 \begin{array}{lll}
 (U_\odot,V_\odot)=(6.53,7.27)\pm(0.24,0.31)~\hbox{km s$^{-1}$},\\
      \Omega_0 =~29.70\pm0.11~\hbox{km s$^{-1}$ kpc$^{-1}$},\\
  \Omega^{'}_0 =-4.035\pm0.031~\hbox{km s$^{-1}$ kpc$^{-2}$},\\
 \Omega^{''}_0 =~0.620\pm0.014~\hbox{km s$^{-1}$ kpc$^{-3}$},
 \end{array}
 \end{equation}
where the error per unit weight is $\sigma_0=12.33$ km s$^{-1}$,
the Oort constants are $A=16.14\pm0.13$ km s$^{-1}$ kpc$^{-1}$ and
$B=-13.56\pm0.17$ km s$^{-1}$ kpc$^{-1}$, and the linear rotation
velocity of the Galaxy at the solar distance is $V_0=237.6\pm4.5$
km s$^{-1}$.

The parameters (12) should be compared with those given in the
last column of Table~2, because both these solutions were obtained
for identical constraints. Such a comparison shows that in the
case where only the proper motions of OB stars (solution (12)) are
used, the kinematic parameters are determined with smaller errors
compared to the results of the simultaneous solution (Table 2).

 \subsection*{OB Stars with Line-of-Sight Velocities}
In this section we analyze the OB stars for which complete
information is available. Thus, there are measurements of their
trigonometric parallaxes, proper motions, and line-of-sight
velocities. Such a sample contains more than 2000 OB stars. For
each such star we can calculate the velocities $U,V,W$ as well as
$V_R$ and $V_{circ}.$ The results of our analysis of these stars
are presented in Figs. 1--4.

Previously (Tables 1 and 2), Eqs. (1) and (2)were solved in
several iterations using the $3\sigma$ criterion to eliminate the
stars with large residuals. Now, when selecting the OB stars, we
use the following additional constraints to improve the quality of
the space velocities (due to the low quality of the series of
line-of-sight velocities for our OB stars):
 \begin{equation}
 \label{cut-2}
 \begin{array}{lll}
 \sqrt{U^2+V^2+W^2}<60~\hbox{km s$^{-1}$},\\
         \qquad |W|<40~\hbox{km s$^{-1}$},
 \end{array}
 \end{equation}
where the rotation curve (12) was taken into account when
calculating the velocities $U,V,$ and $W.$

Figure 1 shows the distribution of 2023 OB stars with relative
parallax errors no more than 30\% on the Galactic $XY$ plane. The
distances to them were calculated by applying the correction
$\Delta\pi=0.050$ mas to the original trigonometric parallaxes.
The Roman numerals in the figure number the following spiral arm
segments: Scutum (I), Carina–Sagittarius (II), Perseus (III), and
the Outer Arm (IV).

Figure 2 plots the circular velocities of 2023 OB stars against
the Galactocentric distance and presents the Galactic rotation
curve constructed according to the solution (12).

Based on the deviation from the Galactic rotation curve (12), we
calculated the residual circular velocities $\Delta V_{circ}$ for
this sample of OB stars. Next, based on the series of their radial
($V_R$) and residual tangential ($\Delta V_{circ}$) velocities, we
found the parameters of the Galactic spiral density wave by
applying a periodogram analysis.

Figure 3 shows the power spectra for the velocities of OB stars.
It is clearly seen from this figure that the peaks of the
distribution lie almost at the same $\lambda$. Indeed, the
perturbation wavelengths are $\lambda_R=2.1\pm0.5$ kpc (for
checking, we calculate the pitch angle $i=-9\pm2^\circ$ for the
adopted four-armed spiral pattern, $m=4,$ from Eq. (8)) and
$\lambda_\theta=1.9\pm0.5$ kpc ($i=-9\pm2^\circ$). The amplitudes
of the radial and tangential velocity perturbations are
$f_R=5.1\pm1.2$ km s$^{-1}$ and $f_\theta=4.4\pm1.4$ km s$^{-1}$,
respectively.

Figure 4 presents the radial and residual tangential velocities of
OB stars. The corresponding periodic curves constructed with the
parameters found as a result of our spectral analysis are shown.
It is clearly seen that these curves, in Figs. 4a and 4b, run with
a phase shift of approximately $90^\circ$. We measure the Sun's
phase in the spiral density wave $\chi_\odot$ from the
Carina-Sagittarius arm ($R\sim7$ kpc); in our case, its value
turned out to be $-178\pm12^\circ$.

 \section*{DISCUSSION}
Based on 130 masers with measured VLBI trigonometric parallaxes,
Rastorguev et al. (2017) found the solar velocity components
$(U_\odot,V_\odot)=(11.40,17.23)\pm(1.33,1.09)$ km s$^{-1}$ and
the following parameters of the Galactic rotation curve:
 $\Omega_0=28.93\pm0.53$ km s$^{-1}$ kpc$^{-1}$,
 $\Omega^{'}_0=-3.96\pm0.07$ km s$^{-1}$ kpc$^{-2}$,
 $\Omega^{''}_0=0.87\pm0.03$ km s$^{-1}$ kpc$^{-3}$, and
  $V_0=243\pm10$ km s$^{-1}$ (for $R_0=8.40\pm0.12$ kpc found).

Based on a sample of 495 OB stars with proper motions from the
Gaia DR2 catalogue, Bobylev and Bajkova (2018b) found the
following kinematic parameters:
 $(U,V,W)_\odot=(8.16,11.19,8.55)\pm(0.48,0.56,0.48)$~km s$^{-1}$,
      $\Omega_0=28.92\pm0.39$~km s$^{-1}$ kpc$^{-1}$,
  $\Omega^{'}_0=-4.087\pm0.083$~km s$^{-1}$ kpc$^{-2}$ and
 $\Omega^{''}_0=0.703\pm0.067$~km s$^{-1}$ kpc$^{-3}$,
where $V_0=231\pm5$ km s$^{-1}$ (for the adopted $R_0=8.0\pm0.15$
kpc). Note that when seeking the solution (12), in this paper we
used an order of magnitude more OB stars.

Here, we have excellent agreement in the parameters found, with
the errors of the parameters being determined in the solution (12)
being very small. In this sense, the solution (12) presently gives
estimates of the parameters $\Omega_0,$ $\Omega^{'}_0$ and
$\Omega^{''}_0$ that are among the best ones.

Note also the parameters found by Bobylev and Bajkova (2019) based
on a sample of 326 young ($log t<8$) OSCs with proper motions and
distances from the Gaia DR2 catalogue:
 $(U,V,W)_\odot=(8.53,11.22,7.83)\pm(0.38,0.46,0.32)$~km s$^{-1}$,
      $\Omega_0=28.71\pm0.22$~km s$^{-1}$ kpc$^{-1}$,
  $\Omega^{'}_0=-4.100\pm0.058$~km s$^{-1}$ kpc$^{-2}$ and
 $\Omega^{''}_0=0.736\pm0.033$~km s$^{-1}$ kpc$^{-3}$.

Parameters of the spiral density wave. The mean pitch angle of the
global four-armed spiral pattern in our Galaxy
$i=-13.6\pm0.4^\circ$. is given in the review of Vall\'ee (1917b).
Then, at $m=4$ and $R_0=8$ kpc $\lambda=3$ kpc follows from Eq.
(8). It can be seen that an analysis of our sample of OB stars
gives a smaller $\lambda$ and, accordingly, a smaller pitch angle
$|i|:9^\circ-10^\circ$.

Having analyzed the spatial distribution of a large sample of
classical Cepheids, Dambis et al. (2015) estimated the pitch angle
of the spiral pattern, $i=-9.5^\circ\pm0.1^\circ$, and the Sun's
phase, $\chi_\odot=-121^\circ\pm3^\circ$, for the four-armed
spiral pattern.

On the other hand, having analyzed maser sources with VLBI
parallaxes, Rastorguev et al. (2017) found
$i=-10.4^\circ\pm0.3^\circ$ and
$\chi_\odot=-125^\circ\pm10^\circ,$ which is in good agreement
with our results.

According to the model estimates by Burton (1971), the amplitudes
of the velocity perturbations from a density wave ($f_R,
f_\theta$) depend on R. Both these velocity perturbations have a
fairly broad maximum in the region $0.5R_0<R<0.9R_0$, reaching
$\approx8$ km s$^{-1}$ ($f_R$ is everywhere larger than $f_\theta$
by $\approx1$ km s$^{-1}$); the velocity perturbations are about 4
km s$^{-1}$ near $0.2R_0$ and decrease to 2.5 km s$^{-1}$ in the
vicinity of $1.3R_0.$

An analysis of the present-day data shows that in a wide vicinity
of $R_0,$ $f_R$ and $f_\theta$ are typically 4--9 km s$^{-1}$ from
masers (Rastorguev et al. 2017), OB stars (Bobylev and Bajkova
2018b), or Cepheids (Bobylev and Bajkova 2012). Note also the new
values of $f_R=4.6\pm0.7$ km s$^{-1}$ and $f_\theta=1.1\pm0.4$ km
s$^{-1}$ obtained recently by Loktin and Popova (2019) from an
analysis of the present-day data on OSCs. The perturbation
amplitudes found in this paper, particularly $f_R,$ are in good
agreement with the results listed above.

When OB stars, young OSCs, or young Cepheids are analyzed, the
absolute value of the phase lies within the range $100-140^\circ$.
As we see, the Sun's phase in the spiral density wave
$-178\pm12^\circ$ found by us differs noticeably from the listed
results of other authors that were obtained by analyzing young
objects.

 \section*{CONCLUSIONS}
We studied the kinematic properties of a large sample of OB stars
($\sim$6000 stars) with proper motions and trigonometric
parallaxes from the GaiaDR2 catalogue and partly also with their
line-of-sight velocities ($\sim$2000 stars). For this purpose, we
used the catalogue of OB stars produced by Xu et al. (2018).

We analyzed the separate and simultaneous solutions of the basic
kinematic equations for various constraints on the relative
trigonometric parallax errors and various corrections of the zero
point of the Gaia DR2 parallaxes, $\Delta\pi$. As a result, we
showed that in the solar neighborhood under consideration (with a
radius of about 4 kpc) the parameters of the smooth Galactic
rotation curve are determined more accurately in the solution
where only the proper motions of OB stars are used (solution
(12)), which was obtained with the involvement of 5335 OB stars.
We showed that there is a noticeable influence of the correction
$\Delta\pi$ when determining the kinematic parameters in the case
where the line-of-sight velocities of OB stars are used.
$\Omega^{''}_0$ is affected most strongly here.

To study the influence of the Galactic spiral density wave, we
used a sample of 2023 OB stars with relative parallax errors no
more than 30\% as well as with known line-of-sight velocities and
proper motions. The amplitudes of the tangential and radial
velocity perturbations produced by the spiral density wave are
$f_\theta=4.4\pm1.4$ km s$^{-1}$ and $f_R=5.1\pm1.2$ km s$^{-1}$,
respectively; the perturbation wavelengths are
$\lambda_\theta=1.9\pm0.5$ kpc and $\lambda_R=2.1\pm0.5$ kpc for
the adopted four-armed spiral pattern. The Sun's phase in the
spiral density wave was found to be
$\chi_\odot=-178^\circ\pm12^\circ$.

 \section*{ACKNOWLEDGMENTS}
We are grateful to the referee for useful remarks that contributed
to an improvement of the paper.

 \section*{FUNDING}
This work was supported in part by the Basic Research Program no.
12 of the Presidium of the Russian Academy of Sciences, the
``Cosmos: Studies of Fundamental Processes and Their
Interrelations'' Subprogram.

 \bigskip \bigskip\medskip{\bf REFERENCES}{\small

1. T. Antoja, A. Helmi, M. Romero-G\'omez, D. Katz, C. Babusiaux,
R. Drimmel, D. W. Evans, F. Figueras, et al., Nature (London,
U.K.) 561, 360 (2018).

2. F. Arenou, X. Luri, C. Babusiaux, C. Fabricius, A. Helmi, T.
Muraveva, A. C. Robin, F. Spoto, et al. (Gaia Collab.), Astron.
Astrophys. 616, 17 (2018).

3. A. T. Bajkova and V. V. Bobylev, Astron. Lett. 38, 549 (2012).

4. V. V. Bobylev and A. T. Bajkova, Mon. Not. R. Astron. Soc. 437,
1549 (2014).

5. V. V. Bobylev and A. T. Bajkova, Mon. Not. R. Astron. Soc. 447,
L50 (2015).

6. V. V. Bobylev and A. T. Bajkova, Astron. Lett. 44, 184 (2018a).

7. V. V. Bobylev and A. T. Bajkova, Astron. Lett. 44, 675 (2018b).

8. V. V. Bobylev, Astron. Lett. 45, 10 (2019).

9. V. V. Bobylev and A. T. Bajkova, Astron. Lett. 45, 109 (2019).

10. A. G. A. Brown, A. Vallenari, T. Prusti, J. de Bruijne, F.
Mignard, R. Drimmel, et al. (Gaia Collab.), Astron. Astrophys.
595, 2 (2016).

11. A. G. A. Brown, A. Vallenari, T. Prusti, J. de Bruijne, C.
Babusiaux, C. A. L. Bailer-Jones, M. Biermann, D. W. Evans, et al.
(Gaia Collab.), Astron. Astrophys. 616, 1 (2018).

12. W. B. Burton, Astron. Astrophys. 10, 76 (1971).

13. J. Byl and M. W. Ovenden, Astrophys. J. 225, 496 (1978).

14. T. Camarillo, M. Varun, M. Tyler, and R. Bharat, Publ. Astron.
Soc. Pacif. 130, 4101 (2018).

15. T. Cantat-Gaudin, C. Jordi, A. Vallenari, A. Bragaglia, L.
Balaguer-Nu\'n\'ez, C. Soubiran, et al., Astron. Astrophys. 618,
93 (2018).

16. A. K. Dambis, L. N. Berdnikov, Yu. N. Efremov, A. Yu. Knyazev,
A. S. Rastorguev, E. V. Glushkova, V. V. Kravtsov, D. G. Turner,
D. J. Majaess, and R. Sefako, Astron. Lett. 41, 489 (2015).

17. E. Franciosini, G. G. Sacco, R. D. Jeffries, F. Damiani, V.
Roccatagliata, D. Fedele, and S. Randich, Astron. Astrophys. 616,
12 (2018).

18. Y. M. Georgelin and Y. P. Georgelin, Astron. Astrophys. 49, 57
(1976).

19. R. de Grijs and G. Bono, Astrophys. J. Suppl. Ser. 232, 22
(2017).

20. A. Helmi, F. van Leeuwen, P. J. McMillan, D. Massari, T.
Antoja, A. C. Robin, L. Lindegren, U. Bastian, et al. (Gaia
Collab.), Astron. Astrophys. 616, 12 (2018).

21. M. Kounkel, K. Covey, G. Su\'arez, C. Rom\'an-Zu\'niga, J.
Hernandez, K. Stassun, K. O. Jaehnig, E. D. Feigelson, et al.,
Astron. J. 156, 84 (2018).

22. C. C. Lin and F. H. Shu, Astrophys. J. 140, 646 (1964).

23. L. Lindegren, J. Hernandez, A. Bombrun, S. Klioner, U.
Bastian, M. Ramos-Lerate, A. de Torres, H. Steidelmuller, et al.
(Gaia Collab.), Astron. Astrophys. 616, 2 (2018).

24. A. V. Loktin and M. E. Popova, Astrophys. Bull. 74 (2019, in
press).

25. J. Maiz-Apellaniz, Astron. J. 121, 2737 (2001).

26. T. Muraveva, H. E. Delgado, G. Clementini, L. M. Sarro, and A.
Garofalo, Mon. Not. R. Astron. Soc. 481, 1195 (2018).

27. M. H. Pinsonneault, Y. P. Elsworth, J. Tayar, A. Serenelli, D.
Stello, J. Zinn, S. Mathur, R. Garcia, et al., Astrophys. J.
Suppl. Ser. 239, 32 (2018).

28. A. S. Rastorguev, M. V. Zabolotskikh, A. K. Dambis, N. D.
Utkin, V. V. Bobylev, and A. T. Bajkova, Astrophys. Bull. 72, 122
(2017).

29. A. G. Riess, S. Casertano, W. Yuan, L. Macri, B. Bucciarelli,
M. G. Lattanzi, J. W. MacKenty, J. B. Bowers, et al., Astrophys.
J. 861, 126 (2018).

30. V. Roccatagliata, G. G. Sacco, E. Franciosini, and S. Randich,
Astron. Astrophys. 617, L4 (2018).

31. D. Russeil, Astron. Astrophys. 397, 133 (2003).

32. C. Soubiran, T. Cantat-Gaudin, M. Romero-Gomez, L.
Casamiquela, C. Jordi, A. Vallenari, T. Antoja, L.
Balaguer-Nu\'nez, et al., Astron. Astrophys. 619, 155 (2018).

33. K. G. Stassun and G. Torres, Astrophys. J. 862, 61 (2018).

34. J. P. Vall\'ee, Astrophys. Space Sci. 362, 79 (2017a).

35. J. P. Vall\'ee, New Astron. Rev. 79, 49 (2017b).

36. Y. Xu, S. B. Bian, M. J. Reid, J. J. Li, B. Zhang, Q. Z. Yan,
T. M. Dame, K. M. Menten, et al., Astron. Astrophys. 616, L15
(2018).

37. L. N. Yalyalieva, A. A. Chemel’, E. V. Glushkova, A. K.
Dambis, and A. D. Klinichev, Astrophys. Bull. 73, 335 (2018).

38. M. V. Zabolotskikh, A. S. Rastorguev, and A. K. Dambis,
Astron. Lett. 28, 454 (2002).

39. E. Zari, H. Hashemi, A. G. A. Brown, K. Jardine, and P. T. de
Zeeuw, Astron. Astrophys. 620, 172 (2018).

40. J. C. Zinn, M. H. Pinsonneault, D. Huber, and D. Stello,
arXiv: 1805.02650 (2018).

  }
  \end{document}